# Role of Secondary Attributes to Boost the Prediction Accuracy of Students' Employability Via Data Mining

Pooja Thakar, Research Scholar
Department of Computer Science
Banasthali University
Jaipur, Rajasthan, India

Prof. Dr. Anil Mehta, Professor
Department of Management
University of Rajasthan
Jaipur, Rajasthan, India

Dr. Manisha, Associate Professor
Department of Computer Science
Banasthali University
Jaipur, Rajasthan, India

*Abstract*—**Data Mining is best-known for its analytical and prediction capabilities. It is used in several areas such as fraud detection, predicting client behavior, money market behavior, bankruptcy prediction. It can also help in establishing an educational ecosystem, which discovers useful knowledge, and assist educators to take proactive decisions to boost student performance and employability.**

**This paper presents an empirical study that compares varied classification algorithms on two datasets of MCA (Masters in Computer Applications) students collected from various affiliated colleges of a reputed state university in India. One dataset includes only primary attributes, whereas other dataset is feeded with secondary psychometric attributes in it. The results showcase that solely primary academic attributes don't lead to smart prediction accuracy of students' employability, once they square measure within the initial year of their education. The study analyzes and stresses the role of secondary psychometric attributes for better prediction accuracy and analysis of students' performance. Timely prediction and analysis of students' performance can help Management, Teachers and Students to work on their gray areas for better results and employment opportunities.**

## I. INTRODUCTION

Every year sizable amount of graduates and postgraduates from numerous professional institutes competes in the job market for good employment opportunities. Conversely, the world economy isn't generating enough employment for this young unemployed populace. To boost the possibility of obtaining the right job that matches the qualification and training of these students, institutes not only ought to add to their academic qualification but should equip them with essential employability skills.

Employability skills are necessary across all areas and kinds of jobs. These skills don't seem to be only important to employees, but employers also seek them in candidates before recruitment. Thus, education system should promote a course of study that emphasizes and nourishes the development of employability skills along with fundamental scholastic skills.

An educational institute contains a large number of student records that remains unutilized. This data can be utilized for the betterment of students if analyzed well. The data can be mined and pruned to guide the students at the right time. Customized guidance to every student in the very first year of their professional college education can give them a rational chance to improve themselves and fetch better employment.

In the higher education system, MCA (Masters in Computer Applications) is a professional degree course that provides theoretical and practical knowledge of Computers and its Applications and makes students ready for IT (Information Technology) and ITeS (Information Technology Enabled Services) Companies. In the last year of the course, every student aspire and compete to obtain a decent job before they pass out. The prediction of students' employability factor and timely steerage by educators can greatly facilitate in rising students efforts in right direction. At the same time better placed and high performer students will bring laurels to the institute reciprocally.

A number of researches have been done to predict students performance, but most of them include only primary academic attributes for prediction purposes. Whereas, many researches have evidently shown positive association between non-academic factors and employability factors. The present study analyzes the role of various factors in improving the prediction accuracy of employability. We applied various data mining classification techniques on student datasets; first with only primary attributes and then one more by adding secondary psychometric attributes to it. Comparative analysis is done by applying classification algorithms in Weka Tool to point out the impact of secondary psychometric attributes on prediction accuracy.

Further, this paper is structured as follows: Section II presents literature review in educational data mining for employability factors prediction and analysis. Section III describes data mining classification techniques used in the study Section IV presents the prediction process of data mining Section V demonstrates and analyzes the results in the form of comparative table and charts. Section VI concludes with an outline and a view on future work.

## II. LITERATURE REVIEW

Data mining has spread its wings in the sector of education very well and lots of work have been done to explore the correlation among attributes, predicting academic performance, finding best mining technique for performance monitoring.

In year 2014, Emerald Group Publishing Limited published a paper stating that emotional intelligence, self management, work and life experiences are necessary factors for Employability Development Profile [1]. Another paper published within the same publishing house and described that employability is joined with competences and tendencies [2]. Cairns, Gueni, Fhima, David and Khelifa analyzed employees' profiles and found positive correlation in employees' jobs, assignments and history [3]. Potgieter & Coetzee, revealed a number of important relationships between personality and employability [4]. David, Hamilton, Riley and Mark disclosed that highest weight is given to soft skills by employers [5]. Denise Jackson and Elaine Chapman in 2012 steered prominent skill gap between professional institutes and corporate [6]. V. K. Gokuladas in his first paper reflected that graduates ought to possess special skills beyond basic academic education [7]. In his next paper he showcased that GPA and proficiency in English language as important factors for employability [8]. Bangsuk Jantawan and Cheng-Fa Tsai, designed a model for prediction of the employees' performance using data mining techniques [9].

Bhardwaj and Pal applied Bayesian classification and found that factors like living location, medium of teaching, mother's qualification, family income and status are highly correlated with academic performance of students [10]. Tongshan Chang, & Ed. D experimented and provided evidences that data mining is an effective technology for college recruitment [11]. Hijazi and Naqvi conjointly found that the factors like mother's education and family income are highly correlated with the student's academic performance [12]. Khan implemented clustering and found that girls with high socio-economic background perform better in science stream and boys with low socio-economic background are usually better academic achiever [13]. Z. J. Kovacic applied CHAID and CART techniques on students enrolment data and presented two decision trees, which classified successful and unsuccessful students. The accuracy obtained with these techniques was only 59.4 % and 60.5 % respectively [14]. Al-Radaideh, et al predicted the final grade of students using ID3, C4.5, and the Naïve Bayes algorithms and found that Decision Tree provide better prediction than any other model [15]. Sudheep Elayidom , Sumam Mary Idikkula & Joseph Alexander applied data mining to assist students in selecting an appropriate branch as per personal skill set for better placement later [16].

These studies reveal great potential of data mining in education sector. More work is needed to establish it as a customized guiding tool for students. With continued research, it'll be ready to support student community very well in near future.

## III. DATA MINING AND CLASSIFICATION

Data mining is often divided into predictive and descriptive techniques. Predictive data mining analyses the data and help in building models, which tries to predict the behaviour of novice instance. The technique of classification belongs to this group and is widely used for predictive modeling. Classification is a supervised learning approach in which data is classified into known classes. Classification rules are identified from training data and are tested for the remainder of data.

The classification accuracy of an algorithm is majorly dependent on the type of dataset rather than chosen algorithm itself. The most important characteristics of a dataset are its predictors, number of classes and number of instances [17]. The used datasets of MCA students do not have much impact of classes, because it has only two output classes namely "employed" and "unemployed". It's binary in nature, therefore works well for most of the algorithms. The number of instances have very little role on the accuracy of classification, the quality of instances matter more [17]. The quality of attributes will offer more information and is a vital factor [17].

Information gain can offer a very good knowledge about the quality of attributes. To find out the quality of attributes in present data sets, information gain is calculated in WEKA Tool for all the attributes. Thereafter, they are stratified from best (Rank 1) to worst (Rank 29) as per the information gain received. The ranks of all attributes present in two datasets are depicted in Fig. 1.

Fig. 1 clearly shows that secondary attributes rank much superior as compared to primary attributes. The attributes like Permanent Address, Attention to Detail, Logical Ability and Score in English are more significant than percentages earned at senior secondary and graduation level. It signifies that only primary academic attributes might not be enough to achieve higher classification accuracy.

To investigate further, two datasets with the same set of instances (214 instances) are taken for comparative analysis between primary and secondary attributes. One dataset contains only primary academic attributes (12) such as percentages earned at secondary, senior secondary and graduation levels, whereas, second dataset includes both primary and secondary psychometric attributes (12 primary and 17 secondary). They're described in Table I.

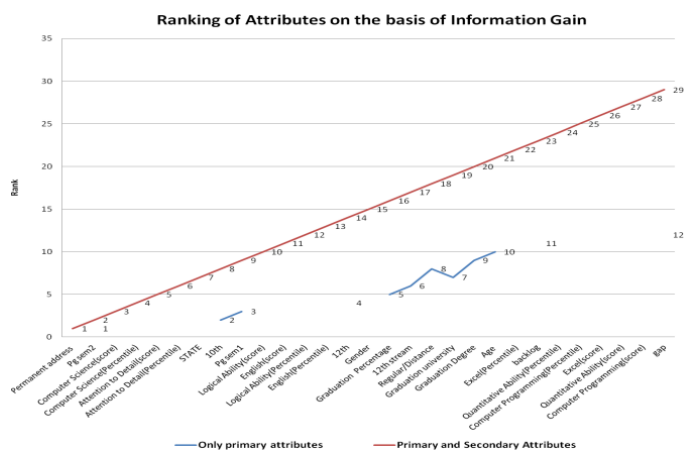

Fig. 1. Graph of attributes ranked as per Information Gain

TABLE I.      LIST OF PRIMARY AND SECONDARY ATTRIBUTES

| Attribute | Values |
|---|---|
| **Primary Attributes** | |
| Age | {<25,>=25} |
| Secondary Percentage | {A,B,C,D,O} |
| Senior Secondary Percentage | {A,B,C,D,O} |
| Stream in Senior Secondary | {Commerce,Science,'Vocational…} |
| Graduation Degree | {CS,'NON CS'} |
| Graduation % | {A,B,C,D,O} |
| Type (Regular/Distance) | {Regular,Distance} |
| Graduating University | {State,Central,Deemed,Private} |
| Post Graduate Sem 1 % | {A,B,O,D,C} |
| Post Graduate Sem 2 % | {A,B,O,D,C} |
| No. of Supplimentaries in 1st Year | {Numeric} |
| Gap in years between Graduation and PG | {Numeric} |
| **Secondary Attributes** | |
| Gender | {MALE,FEMALE} |
| Permanent Address | {'EAST DELHI','OUTSIDE DELHI',…} |
| State | {DELHI,RAJASTHAN,MP,BIHAR,….} |
| English | {Numeric} |
| Quantitative Ability | {Numeric} |
| Logical Ability | {Numeric} |
| Attention to Detail | {Numeric} |
| Computer Programming Skills | {Numeric} |
| Computer Science Knowledge | {Numeric} |
| Psychometric Score | {Numeric} |
| English(P) P→Percentile | {Numeric} |
| Quantitative Ability(P) | {Numeric} |
| Logical Ability(P) | {Numeric} |
| Attention to Detail(P) | {Numeric} |
| Computer Programming(P) | {Numeric} |
| Computer Science(P) | {Numeric} |
| Psychometric (P) | {Numeric} |

Thereafter, various types of classifiers are applied on these two datasets to check which sort of attribute set will herald higher classification accuracy. WEKA Tool is used to implement these classifier algorithms. Ten most widely used base classification algorithms used in educational data mining are chosen. They're Naïve Bayes from Bayes Category, RBF Network and Multilayer Perceptron from Functions Category, IB1 and IBk from Lazy Category, PART and DTNB from Rules Category, lastly J48, Random Tree and Random Forest from Trees Category.

The Classifies are described as below:

Naïve Bayes: This classifier is predicated on Bayes' theorem and believes in individual possibilities of every attribute pair. It is simple to build and easy to understand, withal provides excellent classification results.

RBFNetwork: Radial Basis Function (RBF) networks have proven to be valuable neural network. These are feed-forward networks, which are trained with supervised training algorithm.

The algorithm generally trains in no time and is less susceptible to issues with non-stationary inputs.

Multilayer Perceptron: MLP algorithm is also widely used neural network algorithm. The input layer is of attributes, classes make output layers, hidden layers are interconnected through several neurons. The back propagation algorithm is applied to optimize the weights. This algorithm suits well for approximating a classification.

IB1: The algorithm uses distance measure to find the training instance closest to the given test instance. Thus predicts class that is same as training instance. If multiple instances are same, the first one found is used.

IBk: This is K-nearest neighbours, an instance based classifier. This can select appropriate value of K based on cross-validation and can also do distance weighting.

PART: This algorithm builds a partial C4.5 decision tree with every iteration and makes the "best" leaf into a rule. It uses separate-and-conquer methodology.

DTNB: This is decision table/naive bayes hybrid classifier. The algorithm uses forward selection search. At every step, selected attributes are modeled by naive Bayes. The rest are modeled by decision table. The attribute may be dropped entirely.

J48: J48 decision tree is an implementation of C4.5 algorithm. The tree is structured by training instances and is compatible for dataset with few samples. It doesn't overfit on given dataset.

Random Tree: In this algorithm, K randomly chosen attributes are taken to construct a tree. It doesn't perform any pruning.

Random Forest: This algorithm is used for constructing a forest of random trees.

Robustness of the classifier is usually calculated by applying cross validation on the classifier. During this study, 10-fold cross validation is employed, that split the data set randomly into 10 subsets of equal size. Nine subsets are used as training set and one subset is used as test set. This procedure is performed 10 times to incorporate every subset for test once.

IV.     PREDICTION WITH DATA MINING

Students' employability prediction can facilitate students, academician and management to take proactive actions. This can improve the success percentage of students to get employed in excellent companies. The present study shows that employability can be predicted well, if secondary psychometric parameters are also taken into consideration. Prediction models that embody personal, social, psychological and other environmental variables show better results as compared to models considering only academic parameters.

A. Data Collection

The data set used in this study is obtained from the affiliated colleges of a reputed State University in Delhi, India. Colleges offering three years MCA Degree Course were contacted and complete details of 214 students for the session 2012-2015 were collected.

### B. Data Selection and Transformation

The obtained data set is then divided into two data sets, one with only primary academic attributes (12 attributes) and another with primary and secondary attributes (29 attributes) with the same number of instances (214 instances). Before proceeding for mining, the irrelevant attributes such as name, phone number were removed. Some derived variables like age was added. The data within attributes were also made meaningful by converting them into categories such as marks converted to grades, addresses converted to regions and states. All the predictors are described in Table1.

### C. Implementation of Classification Algorithms

WEKA is an open source mining tool that implements a large collection of machine leaning algorithms. The algorithms used for classification purpose for the present study are Naïve Bayes, RBF Network, Multilayer Perceptron, IB1, IBk, PART, DTNB, J48, Random Tree and Random Forest. The 10-fold cross-validation is chosen as an estimation approach to obtain a reasonable idea of accuracy, since there's no separate test data set. This technique divide training set into 10 equal parts, 9 are applied as training set for making machine algorithm learn and 1 part is used as test set. This approach is enforced 10 times on same dataset, where every training set act as test set once.

## V. RESULTS

The performance of ten classification algorithms for predicting students' employability on two datasets (one with only primary attributes and second with both primary and secondary attributes) were experimented upon and results were calculated.

The percentage of correctly classified instances is commonly known as accuracy or sample accuracy of a model. The accuracy percentage of all the classifiers were calculated for both datasets and results are shown in Table II.

The table is plotted as a graph and is delineated as Fig. 2. It depicts the performance of classifiers with two datasets, initial dataset with only primary attributes and second dataset with both primary and secondary attributes. The graph clearly shows the improvement in performance of most of the classifiers, when second dataset is chosen. This further proves that the information gain shown by the secondary attributes earlier (Fig. 1), additionally facilitate in better performance of classifiers. Thus, we can say that secondary attributes play very important role in improving prediction accuracy of classifiers with respect to employability. Along with the results of accuracy, the training and simulation errors with the help of Kappa Statistic, Mean Absolute Error (MAE) Root Mean Squared Error (RMSE), Relative Absolute Error (RAE) and Root Relative Squared Error (RRSE) were calculated. The results of simulation are shown in Table III and Table IV. The percentage differences of RAE% and RRSE% are further plotted in a graph and are depicted in Fig. 3 and Fig. 4. The graphs clearly show that Error Percentages (RAE and RRSE) reduces significantly, when we include secondary psychometric parameters in dataset.

TABLE II.    PREDICTION ACCURACY OF CLASSIFIERS

| Classifier Name | Accuracy % (Only Primary Attributes) | Accuracy % (Primary and Secondary Attributes) |
|---|---|---|
| NaïveBayes | 77.10% | 84.50% |
| RBFNetwork | 83.10% | 85.04% |
| Multilayer Perceptron | 73.80% | 82.71% |
| IB1 | 74.20% | 83.17% |
| IBk | 80.30% | 83.17% |
| PART | 81.30% | 85.90% |
| DTNB | 78.90% | 86.40% |
| J48 | 84.50% | 84.50% |
| Random Tree | 79.90% | 80.37% |
| RandomForest | 79.40% | 85.90% |

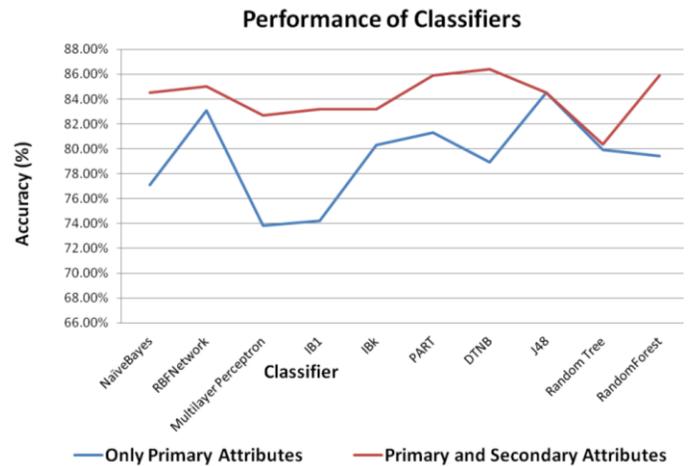

Fig. 2.   Graph depicting Classifiers Performance

TABLE III.    TRAINING AND SIMULATION ERROR PERCENTAGE

| Classifier | RAE (Only Primary Attributes) | RAE (Primary and Secondary Attributes) | RRSE (Only Primary Attributes) | RRSE (Primary and Secondary Attributes) |
|---|---|---|---|---|
| NaïveBayes | 100.41% | 66.40% | 111.04% | 99.07% |
| RBFNetwork | 95.60% | 76.30% | 103.20% | 93.30% |
| Multilayer Perceptron | 96.60% | 67.60% | 126.96% | 102.90% |
| IB1 | 97.50% | 63.80% | 140.30% | 113.50% |
| IBk | 92.09% | 65.10% | 119.80% | 112.90% |
| PART | 98.21% | 84.28% | 101.47% | 92.74% |
| DTNB | 130.50% | 102.20% | 111.90% | 98.07% |
| J48 | 99.06% | 94.94% | 99.99% | 99.10% |
| Random Tree | 92.60% | 86.04% | 121.60% | 114.90% |
| RandomForest | 99.60% | 90.20% | 109.60% | 95.10% |

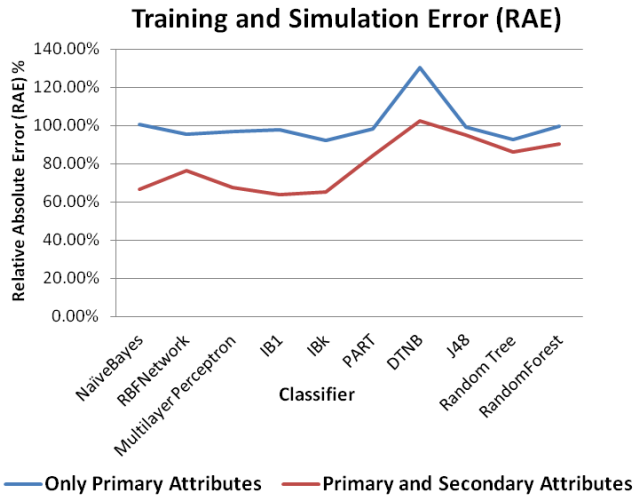

Fig. 3.  Graph of Relative Absolute Error Percentages for classifiers

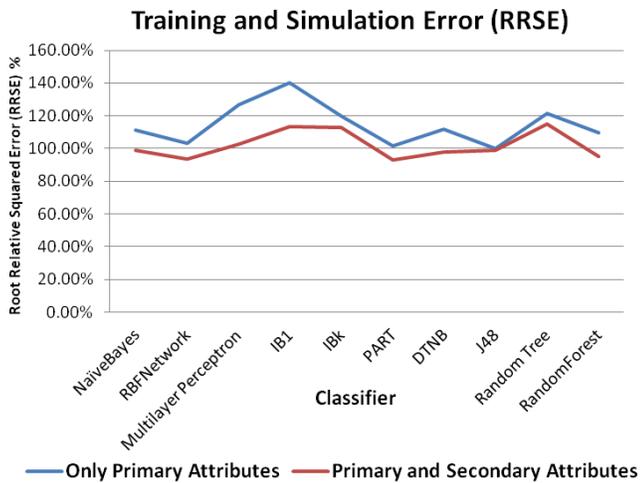

Fig. 4.  Graph of Root Relative Squared Error Percentages for classifiers

Kappa Statistics: Kappa is a normalized value of agreement for chance. It can be described as

K= (P (A)- P(E))/(1-P (E))
Where,

P (A) is percentage agreement and P (E) is chance agreement.

If K =1 than agreement is ideal between the classifier and ground truth.

If K=0, it indicates there's a chance of agreement.

Table IV represents the Kappa Statistics calculated for every classifier used in the current study, once for initial dataset with primary academic attributes only, thenceforth with second dataset, which includes both primary and secondary attributes. Each classifier produces K value greater than 0 i.e. each classifier is doing better than the chance for training set [18], once second dataset is chosen. Fig. 5 additionally depicts the values in graph form.

TABLE IV.    KAPPA STATISTIC OF CLASSIFIERS

| Classifier | Kappa Statistic (Only Primary Attributes) | Kappa Statistic (Primary and Secondary Attributes) |
|---|---|---|
| NaïveBayes | 0.0389 | 0.386 |
| RBFNetwork | -0.0264 | 0.1544 |
| Multilayer Perceptron | 0.0211 | 0.3116 |
| IB1 | -0.0503 | 0.0126 |
| IBk | -0.0002 | 0.0126 |
| PART | 0 | 0.2612 |
| DTNB | -0.0543 | 0.2978 |
| J48 | 0 | 0.0773 |
| Random Tree | 0.0818 | 0.1872 |
| RandomForest | -0.0835 | 0.2072 |

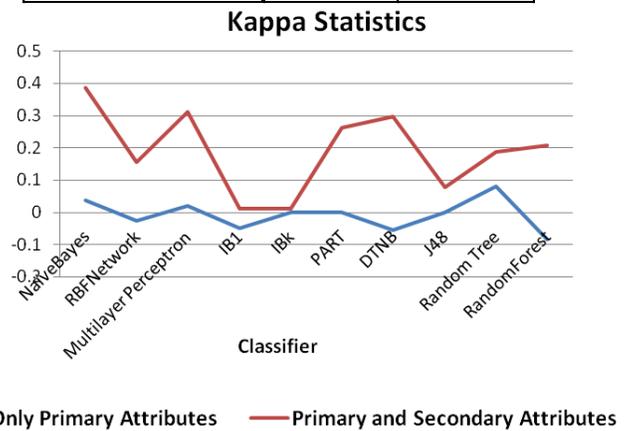

Fig. 5.  Graph of Kappa Statistics Percentages for classifiers

Once prognostic model is formed, it is necessary to ascertain its accuracy. It is generally calculated based on the precision, recall values of classification matrix.

Precision is a fraction of retrieved instances that are relevant. It is calculated as

PRECISION= ( TP)/(TP+FP)

Where,

TP is total number of true positives.

FP is total number of false positives.

Recall is a fraction of relevant instances that are retrieved. It is usually expressed in percentages and is calculated as

RECALL= ( TP)/(TP+FN)

Where,

TP is total number of true positives.

FN is total number of false negatives.

These methods are not very apposite, if dataset is imbalanced [19]. The datasets utilized in the current study is imbalanced with only few instances for "employed" class and large number of instances for "unemployed" class.

Receiver Operating Characteristic (ROC) Curve/Area is suggested to be a better choice of evaluation with such dataset [19].

ROC curves can represent the family of best decision boundaries for relative costs of True Positive (TP) and False Positive (FP).

In ROC curve the X-axis represents

% FP=(False Positive)/(True Negative+False Positive)

and the Y-axis represents

% TP=(True Positive)/(True Positive+False Negative)

The ideal point on the ROC curve is (0,100) that is when all positive examples are classified correctly and no negative examples are misclassified as positive.

Area under the ROC Curve (AUC) is a very useful metric for judging classifier performance. It is independent of the decision criterion selected and prior probabilities. The AUC comparison can ascertain a dominance relationship between classifiers.

Comparison of evaluation measure ROC Area for minority class "employed" is presented in Table V. It is also further depicted as graph in Fig. 6.

The graph (Fig. 6) clearly illustrates the increase in ROC Area values for almost all the classifiers towards 1, when second dataset is chosen as compared to the first dataset with only primary attributes.

This also implies and proves that the performances of learning techniques are highly dependent on the nature of the dataset used.

TABLE V.      PERFORMANCE OF CLASSIFIERS W.R.T ROC AREA

| Classifier | ROC Area (Only Primary Attributes) | ROC Area (Primary and Secondary Attributes) |
|---|---|---|
| Naïve Bayes | 0.619 | 0.868 |
| RBFNetwork | 0.596 | 0.784 |
| Multilayer Perceptron | 0.595 | 0.765 |
| IB1 | 0.476 | 0.504 |
| IBk | 0.566 | 0.545 |
| PART | 0.501 | 0.719 |
| DTNB | 0.602 | 0.701 |
| J48 | 0.466 | 0.635 |
| Random Tree | 0.579 | 0.697 |

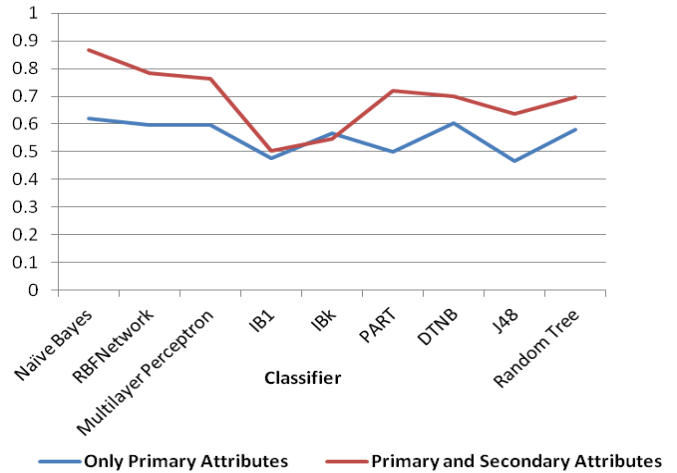

Fig. 6.   Graph of ROC Area for minority class of classifiers

## VI.   CONCLUSION AND FUTURE SCOPE

The results prove that prediction accuracy for students' employability can be enhanced with the inclusion of secondary attributes such as personal, social, psychological and other environmental variables in the dataset. Accuracy percentage shows incredible improvement with all types of classifiers. Error Percentage also reduces remarkably. Kappa Statistics and ROC Area shows great signs of improvement. Hence, proves that secondary psychometric attributes play the essential role in boosting the prediction accuracy of students' employability.

Due to imbalanced datasets, classifiers could not attain high percentage accuracy with low error percentage. The maximum accuracy percentage attained in the study is by DTNB, which is 86.4% with very high error percentages, that crosses the minimal limit. Thus may not be helpful enough to be converted into prediction rules. Some technique is required to handle the problem of the imbalanced dataset.

In future, the dataset can be improved by adding more significant attributes to enhance the accuracy percentage of classifiers with low error percentage. Moreover, some automated technique is also required at the preprocessing stage to identify the best attributes for finest performance of classifiers; which also handles the problem of an imbalanced dataset.